\documentclass[conference]{IEEEtran}
\IEEEoverridecommandlockouts

\usepackage{cite}
\usepackage{amsmath,amssymb,amsfonts}
\usepackage{graphicx}
\usepackage{textcomp}
\usepackage{xcolor}
\usepackage{multirow}
\usepackage{url}
\usepackage{algorithm}
 \usepackage{algpseudocode}
 \usepackage{algorithmicx}
 \algdef{SE}[DOWHILE]{Do}{doWhile}{\algorithmicdo}[1]{\algorithmicwhile\ #1}%
 
\def\BibTeX{{\rm B\kern-.05em{\sc i\kern-.025em b}\kern-.08em
    T\kern-.1667em\lower.7ex\hbox{E}\kern-.125emX}}
\begin{document}

\title{A Synergy between On- and Off-Chip Data Reuse for GPU-based Out-of-Core Stencil Computation}

\author{\IEEEauthorblockN{Jingcheng Shen}
\IEEEauthorblockA{\textit{Chongqing University of} \\
\textit{Posts and Telecommunications}\\
Chongqing, China\\
shenjc@cqupt.edu.cn}
\and
\IEEEauthorblockN{Linbo Long}
\IEEEauthorblockA{\textit{Chongqing University of} \\
\textit{Posts and Telecommunications}\\
Chongqing, China\\
longlb@cqupt.edu.cn}
\and
\IEEEauthorblockN{Jun Zhang}
\IEEEauthorblockA{\textit{Arizona State University} \\
Tempe, AZ, USA \\
jeffzhang@asu.edu}
\and
\IEEEauthorblockN{Weiqi Shen}
\IEEEauthorblockA{\textit{AECC Commercial Aircraft Engine CO., LTD} \\
Shanghai, China \\
weiqishen1994@gmail.com}
\and
\IEEEauthorblockN{Masao Okita}
\IEEEauthorblockA{
\textit{Osaka University}\\
Osaka, Japan \\
okita@ist.osaka-u.ac.jp}
\and
\IEEEauthorblockN{Fumihiko Ino}
\IEEEauthorblockA{
\textit{Osaka University}\\
Osaka, Japan \\
ino@ist.osaka-u.ac.jp}
}

\maketitle

\begin{abstract}
Stencil computation is an extensively-utilized class of scientific-computing applications that can be efficiently accelerated by graphics processing units (GPUs).
Out-of-core approaches enable a GPU to handle large stencil codes whose data size is beyond the memory capacity of the GPU.
However, current research on out-of-core stencil computation primarily focus on minimizing the amount of data transferred between the CPU and GPU. Few studies consider simultaneously optimizing data transfer and kernel execution. 
To fill the research gap, this work presents a synergy between on- and off-chip data reuse for out-of-core stencil codes, termed SO2DR. 
First, overlapping regions between data chunks are shared in the off-chip memory to eliminate redundant CPU-GPU data transfer.
Secondly, redundant computation at the off-chip memory level is intentionally introduced to decouple kernel execution from region sharing, hence enabling data reuse in the on-chip memory.
Experimental results demonstrate that SO2DR significantly enhances the kernel-execution performance while reducing the CPU-GPU data-transfer time. 
Specifically, SO2DR achieves average speedups of $2.78\times$ and $1.14\times$ for five stencil benchmarks, compared to an out-of-core stencil code which is free of redundant transfer and computation, and an in-core stencil code which is free of data transfer, respectively. 
\end{abstract}


\section{Introduction}
Stencil computation is an important class of applications in computer science and scientific computing, occurring in a wide range of fields such as geophysics simulations~\cite{serpa17padw,farres19eage}, fluid dynamics~\cite{wichmann2019practical,huckelheim2019automatic}, image processing~\cite{ikeda2014efficient,tabik2018tuning}. 
Stencil computation updates every element in an array based on the values of its neighboring elements according to a computing template, i.e., stencil.
Moreover, the updating operations on different elements are mutually independent, hence stencil computation is an embarrassingly parallel scenario to leverage accelerators such as graphics processing units (GPUs).
A GPU has thousands of cores and its memory bandwidth is 5$-$10$\times$ as high as that of a CPU, thus extensively utilized in accelerating compute- and memory-intensive applications~\cite{ino2014parallel,mitani2016parallelizing,shen2019gpu}.
Nonetheless, GPUs possess a relatively limited device-memory capacity, typically in the range of several dozen GBs. 
Consequently, a GPU fails to directly accelerate a large-scale stencil code that operates on datasets exceeding the memory capacity. 

Numerous studies have successfully ported large-scale stencil codes to GPUs by utilizing out-of-core methodologies~\cite{hou2017gpu,shimokawabe2017stencil,sourouri2017panda}. For data exceeding GPU memory, the out-of-core approach first decomposes data into moderately-sized chunks, each can fit into the GPU memory. Subsequently, these chunks are systematically streamed to and from the GPU until the entirety of the data is processed.
Nevertheless, the performance of the out-of-core approach is often reported to be limited by the overhead of data transfer between the CPU and GPU. 
Techniques as temporal blocking (TB)~\cite{miki2019pacc,perepelkina2021extending} and region sharing~\cite{jin13cluster,reguly17mchpc,shen20ieice,shen23supe} are utilized to enhance on-GPU data reuse and eliminate redundant data transfer.
Nevertheless, existing research concerning out-of-core stencil computation primarily concentrates on reducing the amount of data transferred between the CPU and GPU, ignoring simultaneously leveraging data reuse in the on-chip memory (e.g., register files and scratchpad memory) to enhance the performance of kernels (computational code executed on the GPU). 
Their focus is well-founded, given the widespread assumption that the bottleneck in out-of-core stencil computation resides within the CPU-GPU data transfer.

However, we have noted that this assumption does not invariably hold true. 
The bottleneck of out-of-core stencil computation can transition from the CPU-GPU data transfer to the execution of GPU kernels, contingent upon shifts in the environment or alterations in the configuration of the stencil computation process. 
In such cases, we should prioritize the reduction of kernel-execution time over that of data transfer time. 
To fulfill the goal, some studies~\cite{rawat2018register,rawat2019optimizing,matsumura2020an5d} intelligently utilize the on-chip memory during kernel execution.
Among these studies, AN5D~\cite{matsumura2020an5d} is a prominent framework to generate high-performance kernels for GPU-based stencil codes. 
AN5D-generated stencil kernels exploit on-chip memory of the GPU for data reuse, thus reducing data accesses to device memory. In doing so, AN5D-generated kernels benefit from a significantly reduced execution time.
Nonetheless, these studies are originally designed for in-core stencil codes, assuming that the GPU hold the entirety of data required by the stencil computation. Such an assumption fails to hold for the out-of-core stencil computation because the GPU cannot store the entirety of data in its limited memory.

To bridge this research gap, we propose SO2DR (pronounced as ``solder''), a \textbf{S}ynergy between \textbf{O}n- and \textbf{O}ff-chip \textbf{D}ata \textbf{R}euse for out-of-core stencil computation on a GPU. 
SO2DR can not only eliminate the redundant CPU-GPU data transfer by enabling data reuse between chunks residing on the off-chip (i.e., device) memory, but also notably enhancing the kernel execution time by exploiting on-chip memory to allow data reuse between multiple time steps.
Furthermore, SO2DR introduces redundant computations to resolve conflicts between TB in the off-chip memory and that in the on-chip memory.
The contributions of this work are summarized as follows.
\begin{itemize}
    \item A novel method named SO2DR that harnesses a synergy between on- and off-chip data reuse is presented, which reduces kernel execution time as well as data transfer time for GPU-based out-of-core stencil computation.
    \item An in-depth discussion on addressing the performance bottleneck within GPU-based out-of-core stencil codes is given to describe the motivation of this work and provide a valuable perspective to guide further research.
    \item Comprehensive experiments are conducted to evaluate the effectiveness of SO2DR. The experimental results are meticulously analyzed to rationalize the performance achievement and discuss the potential of SO2DR.   
\end{itemize}

The remainder of this paper is organized as follows. Background and motivation of this work are described in Section~\ref{sec:bgd} and Section~\ref{sec:motiv}, respectively. 
SO2DR that employs the synergy between on- and off-chip data reuse is presented in Section~\ref{sec:prop}.
Experimental results are presented and analyzed in Section~\ref{sec:expr}. 
Section~\ref{sec:rel} provides a literature review of research related to accelerating large-scale stencil codes on GPUs.
Finally, Section~\ref{sec:conc} concludes this paper and suggests future research directions.

\begin{figure}[tb!]
    \centering
    \includegraphics[width=\linewidth]{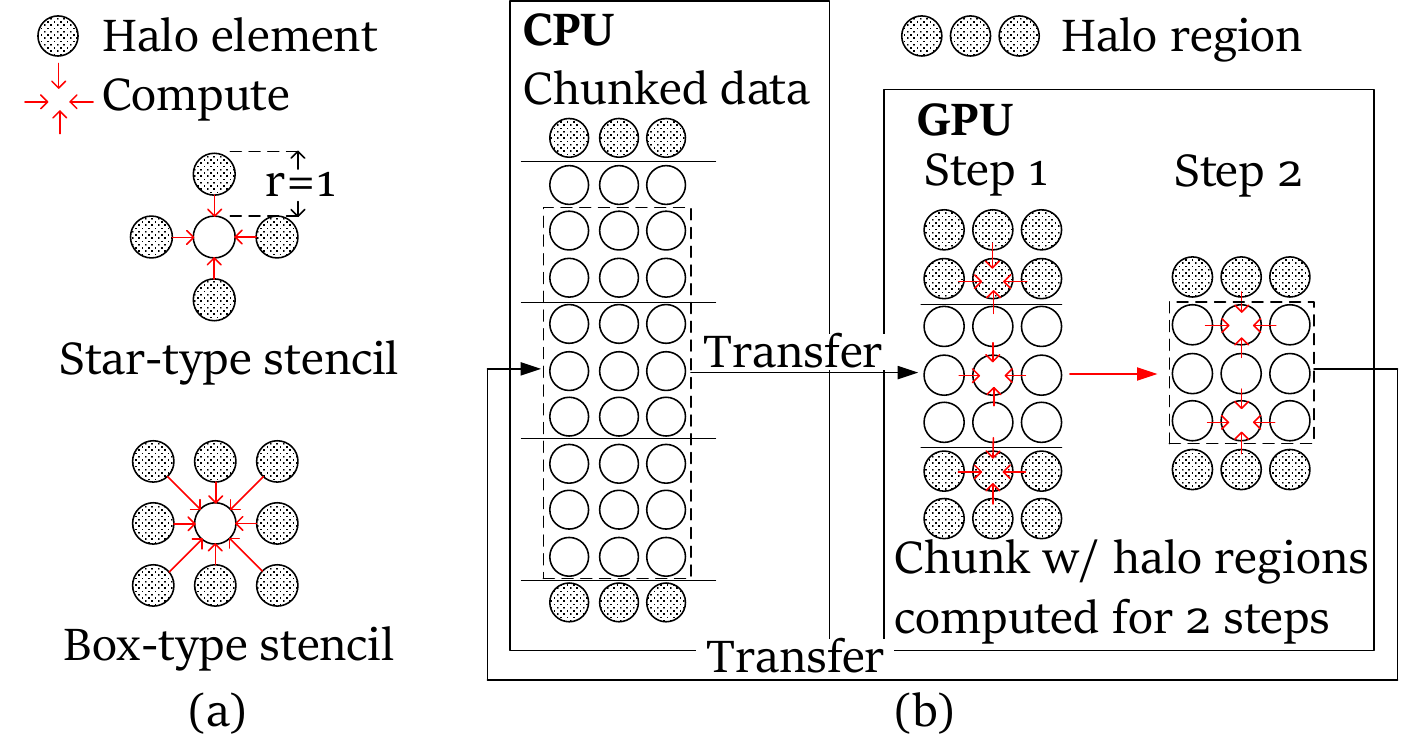}
    \caption{Overview of stencil computation. (a): Star- and box-type stencils. (b):~Out-of-core stencil computation. Note that computing for $k$ steps on the GPU requires $k$ pieces of halo regions to be transferred in addition to the chunk. Moreover, ``r'' denotes the stencil radius.}
    \label{fig:stencil}
\end{figure}
\section{Background}
\label{sec:bgd}
This section presents the attributes of GPU-based out-of-core stencil computation and the typical technique employed for its optimization. 
\subsection{Stencil Computation}
Stencil computation involves an iterative data processing approach that updates each element of the given array based on predetermined patterns known as stencils. Within a stencil, the value of an element is computed using values of the neighboring (i.e., halo) elements (Fig.~\ref{fig:stencil}a).

Out-of-core approaches address the challenge posed by large-scale stencil codes wherein the data size exceeds the capacity of the GPU memory. 
In out-of-core stencil computation, the initial dataset is partitioned into small chunks, each of which can fit in the GPU memory. These chunks are then systematically transferred to the GPU for processing.
Nevertheless, this methodology involves frequent data transfers between the CPU and GPU, consequently introducing a performance bottleneck that impairs the overall performance.

To mitigate the side effect of out-of-core computation, the TB technique is widely leveraged, which will be described in the subsequent section.
\subsection{Temporal Blocking (TB)}
The TB technique is extensively used to reduce the amount of data transfer in hierarchical memory systems. 
For GPU-based out-of-core stencil computation, the TB technique transfers chunks with halo regions that allow a chunk on the GPU to be processed for multiple time steps before being transferred back to the CPU (Fig.~\ref{fig:stencil}b). 
If a stencil code employs $k$ TB steps, it requires only $1/k$ times of CPU-GPU data transfer, compared to the code without TB.

\begin{figure}
    \centering
    \includegraphics[width=\linewidth]{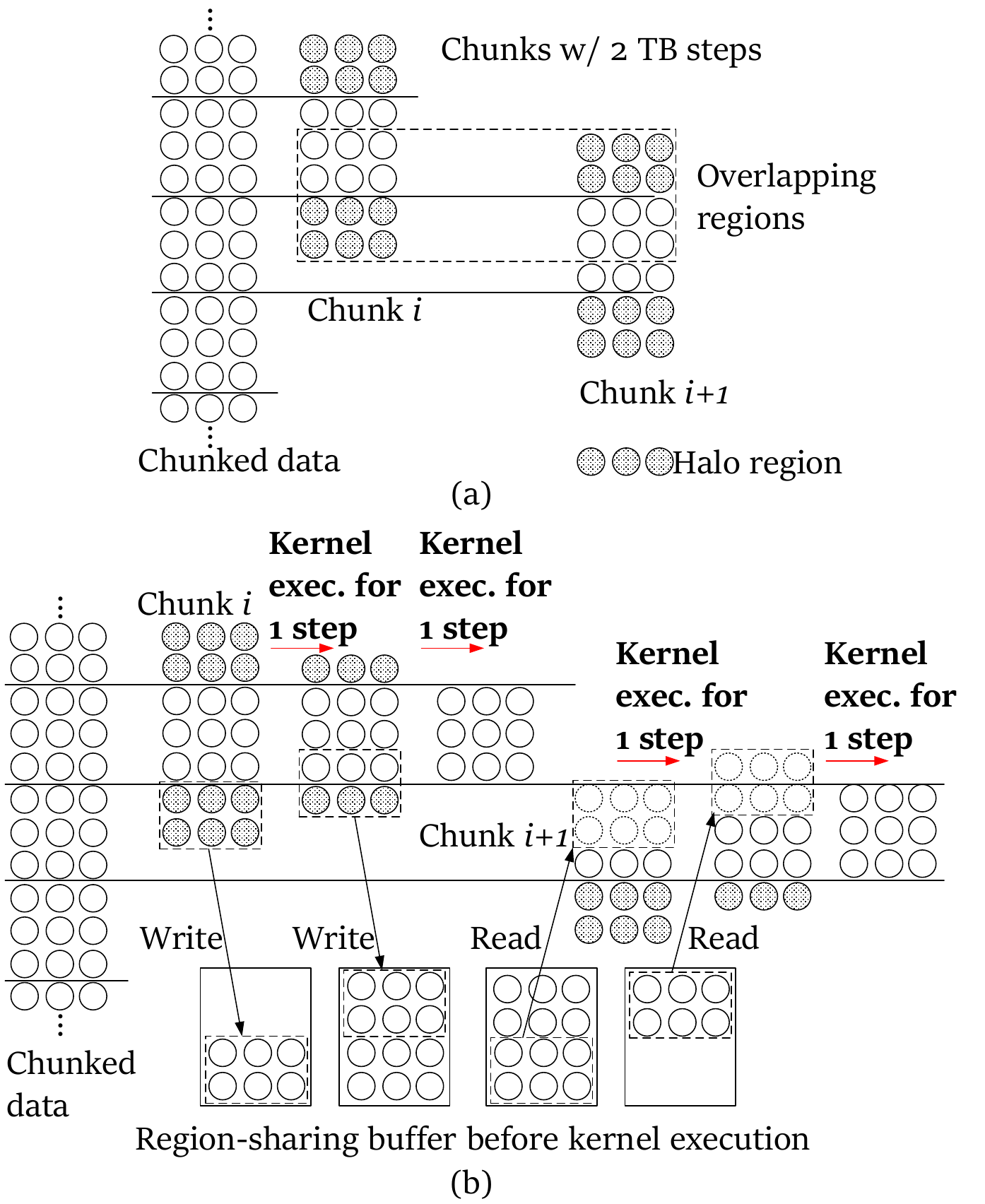}
    \caption{Region-sharing technique. (a): Overlapping regions that can be shared between two adjacent chunks. (b):~Eliminating redundant data transfer by allowing two adjacent chunks to share the data in the overlapped regions on the GPU using a region-sharing buffer. \textit{Note that because sharing intermediate results at off-chip memory level is required between GPU kernels, each GPU kernel can only execute one time step.}}
    \label{fig:share}
\end{figure}

However, increasing the number of TB steps simultaneously expands the halo regions that must be transferred alongside each chunk, significantly burdening the interconnect between the CPU and GPU. A region-sharing technique~\cite{jin13cluster} is thus implemented to eliminate the amount of redundant data transfer.
The concept behind the region-sharing technique stems from the observation that adjacent chunks possess overlapping regions that can be mutually shared (Fig.~\ref{fig:share}a).

In line with the technique, once a chunk has been transferred to the GPU, it shares overlapping regions to the next chunk so that the next chunk can be transferred without the redundant data (Fig.~\ref{fig:share}b).
Precisely, before each time step of GPU computation, the chunk retrieves two shared regions from the region-sharing buffer. The buffer stores intermediate computation results of the prior chunk. Likewise, the chunk writes two shared regions to the region-sharing buffer for the subsequent chunk to reuse.
Effectually, the region-sharing technique not only eliminates the amount of redundant data transfer between the CPU and GPU, but also avoids redundant computation on the GPU.
However, such reuse of intermediate computation results in the off-chip memory prevents GPU kernels from exploiting on-chip data reuse, which will be elaborated on in Section~\ref{sec:motiv}.  

\section{Motivation}
\label{sec:motiv}
\begin{figure}[tb!]
    \centering
    \includegraphics[width=\linewidth]{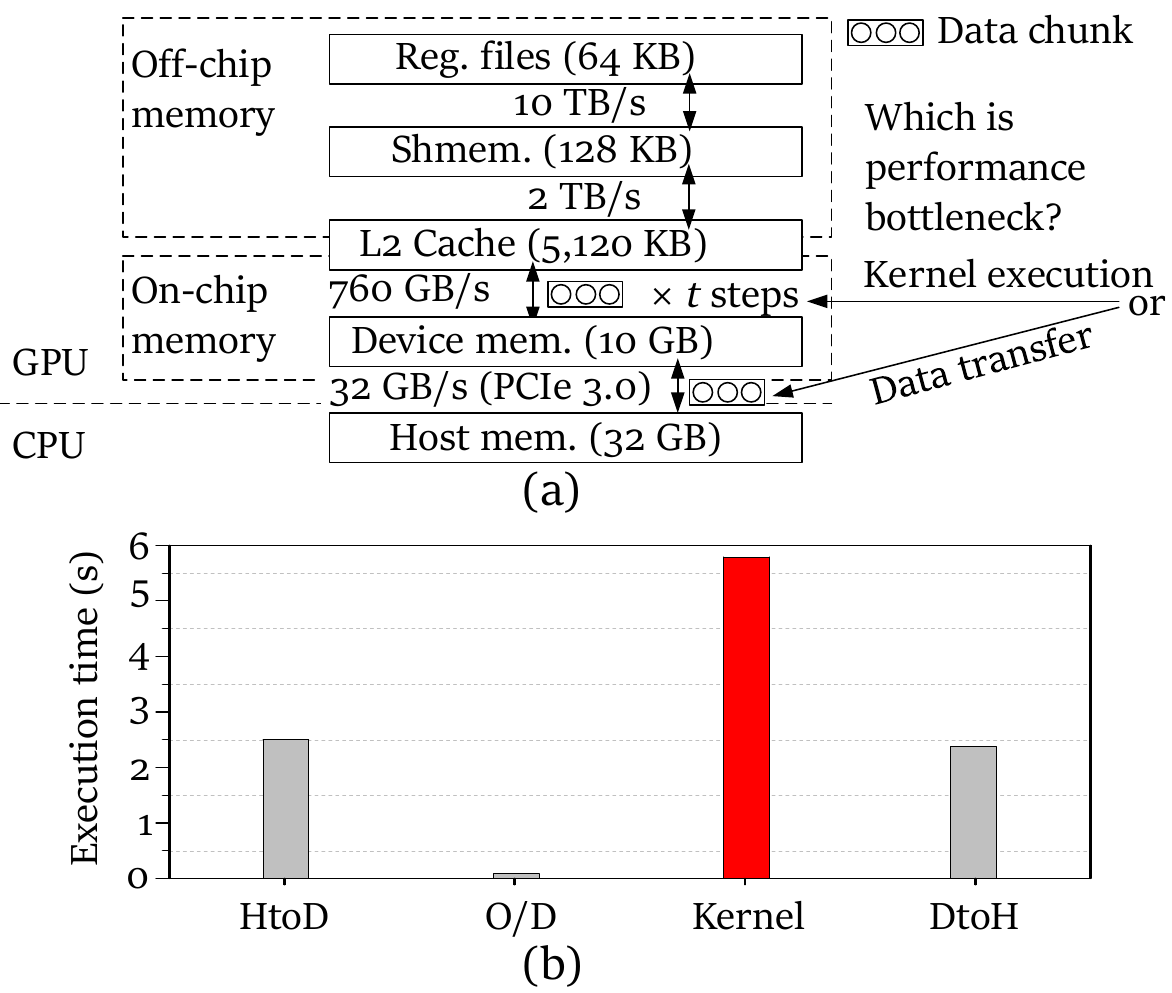}
    \caption{Motivation of this work. (a): Choice of optimization target. (b):~Preliminary results showcasing a kernel-execution bottleneck. Bandwidths and memory sizes here belong to our experimental machine (refer to Table~II). ``Shmem,'' ``HtoD,'' ``O/D,'' and ``DtoH'' denote ``shared memory," ``host-to-device data transfer," ``on-device data copy,'' and ``device-to-host data transfer,'' respectively.}
    \label{fig:motiv}
\end{figure}
Existing research primarily focuses on minimizing data transfers under the assumption that the performance bottleneck of out-of-core stencil computation lies in the CPU-GPU data-transfer overhead.
However, the assumption may cease to hold true if conditions change.
The task to optimize an out-of-core stencil code can be roughly modeled below:
 \begin{eqnarray*}
   \begin{array}{lllll}
    \mathrm{minimize}  & T_{tot} \propto Max(\frac{D_{chk}}{BW_{intc}},\frac{D_{chk}+W_{halo}\times S_{TB}}{BW_{dmem}}\times{S_{TB}}), & 
\\\mathrm{subject~to} & (D_{chk}+W_{halo} \times S_{TB}) \times N_{strm} \leq C_{dmem}.&\\
   \end{array}
 \end{eqnarray*}
Here, $T_{tot}$ represents the total execution time of the stencil code. $D_{chk}$ denotes the size of the data chunk to be transferred to the GPU for processing. $BW_{intc}$ corresponds to the bandwidth of the interconnect (e.g., the PCIe bus) between the CPU and GPU, and $BW_{dmem}$, the bandwidth of the GPU off-chip memory. 
$S_{TB}$ stands for the number of TB steps. $W_{halo}$ represents the size of working space required by each halo region. $N_{strm}$ denotes the number of GPU operation (i.e., CUDA) streams~\cite{cudawww}. 
Lastly, $C_{dmem}$ stands for the capacity of GPU off-chip memory. Note that a GPU can leverage multiple operation streams to overlap data transfer with kernel execution.    

With the elimination of redundant data transfer achieved by the region-sharing technique, the data transfer time is static for a given chunk size and interconnect bandwidth. Contrarily, the kernel execution time increases in conformity with the number of TB steps. Note that the kernel performance is limited by the data movement from the off-chip memory to on-chip memory (which is near the compute cores). For each TB step, data is moved from the off-chip memory to on-chip memory if on-chip data reuse is not employed. 
Figure~\ref{fig:motiv}a depicts the choice of optimization target, whether it be kernel execution or data transfer.

\textit{Therefore, when the number of TB steps is large enough such that the kernel execution time is longer than the CPU-GPU data-transfer time, our attention must pivot towards reducing the kernel-execution time.}
For instance, Figure~\ref{fig:motiv}b illustrates preliminary experimental results to showcase a kernel-execution bottleneck. In this experiment, a box-type stencil code with stencil radius$=$1 ran for a total number of 320 time steps on the GPU. 
The dataset amounts to 11 GB, divided into eight chunks. The number of TB steps is set at 40.
The kernel-execution time was found to be 2.3$\times$ as long as the HtoD data-transfer time, indicating the bottleneck resides in kernel execution.

\textit{On-chip data reuse can effectively reduce the amount of data movement between the off- and on-chip memories, which therefore significantly enhances kernel-execution time.}
Stencil codes generated by the AN5D framework harness on-chip data reuse to allow a GPU kernel to proceed multiple steps, hence conducting TB at the on-chip memory level.
\textit{However, the on-chip data reuse conflicts with the reuse of intermediate results at the off-chip memory level, where the kernel executions are interleaved with reads and writes of intermediate results between time steps (Fig.~\ref{fig:share}b). Consequently, each GPU kernel can only proceed one time step.} 
Given the analysis, this work is aimed at combining the on- and off-chip data reuse to enhance the overall performance of out-of-core stencil codes whose bottleneck resides in kernel execution due to large numbers of TB steps.

\section{Proposed Method}
\label{sec:prop}
This section presents in detail the proposed SO2DR that harnesses a synergy between on- and off-chip data use. First, the synergy can both eliminate redundant CPU-GPU data transfer and enhance GPU kernel execution via reducing data traffic between the on- and off-chip memories. 
Secondly, redundant computation in the off-chip memory is deliberately introduced so as to decouple kernel execution from data sharing at the off-chip memory level, paving the way for GPU kernels to exploit on-chip data reuse.
\subsection{Synergy between On- and Off-Chip Data Reuse}
\begin{algorithm}[t!]    
    \caption{Proposed method that harnesses both on- and off-chip data reuse.}\label{alg:prop}
    \begin{algorithmic}[1]
        \Require{$d$: no. of chunks, $n$: total no. of iterations, $k_{off}$: no. of TB steps in GPU off-chip memory, and $k_{on}$: no. of steps processed w/i a kernel using on-chip data reuse.}
        \Ensure{Updated chunks.}
        \State $N_{t} \leftarrow \frac{n+k_{off}-1}{k_{off}}$
        \State \textbf{for} $t \leftarrow 0;t<N_{t};t++$ \textbf{then}
        \State \hspace*{2mm} $k^{'}_{off} \leftarrow$  \textbf{if} $t=N_{t}-1$ \&\& $n\%{k_{off}}\neq0$ \textbf{then} $n\%{k_{off}}$ \textbf{else} $k_{off}$
        \State \hspace*{2mm} \textbf{for} $i \leftarrow 0;i<d;i++$ \textbf{then}
        \State \hspace*{2mm}\hspace*{2mm} Transfer $i$-th chunk to GPU
        \State \hspace*{2mm}\hspace*{2mm} Read and then write regions from and to region-sharing buffer
        \State \hspace*{2mm}\hspace*{2mm}\textbf{for} $j \leftarrow 0;j<\frac{k^{'}_{off}}{k_{on}};j++$ \textbf{then}
        \State \hspace*{2mm}\hspace*{2mm}\hspace*{2mm} Adjust compute area of $i$-th chunk according to $j$
        \State \hspace*{2mm}\hspace*{2mm}\hspace*{2mm} Apply a $k_{on}$-step kernel to $i$-th chunk
        \State \hspace*{2mm} \hspace*{2mm} \textbf{end for}
        \State \hspace*{2mm}\hspace*{2mm}\textbf{if} $k^{'}_{off}\%k_{on}\neq 0$ \textbf{then}
        \State \hspace*{2mm}\hspace*{2mm}\hspace*{2mm} Adjust compute area of $i$-th chunk according to $j$
        \State \hspace*{2mm}\hspace*{2mm}\hspace*{2mm} Apply a $\{k^{'}_{off}\%k_{on}\}$-step kernel to $i$-th chunk
         \State \hspace*{2mm}\hspace*{2mm}\hspace*{2mm}\textbf{endif}
        \State \hspace*{2mm} \hspace*{2mm}Transfer $i$-th chunk back to CPU
        \State \hspace*{2mm} \textbf{end for}
        \State \textbf{end for}
        \State \textbf{return} chunks
    \end{algorithmic}
\end{algorithm}
The key idea of SO2DR is concise. After sharing overlapping regions with its adjacent counterpart, each chunk is enabled to be processed for multiple time steps without any interruption during kernel execution.
On-chip data reuse can be exploited in this manner exclusively due to the fact that the on-chip memories are solely exposed to collaborative threads within the same kernel.

Algorithm 1 illustrates the general workflow of SO2DR. For a total number of $n$ iterations, each of the $d$ chunks must be transferred to and from the GPU for $(n+k_{off}-1)/k_{off}$ times. Note that if $n$ is indivisible by $k_{off}$, the chunk is processed for the residual number of steps on the GPU when the last time ($T_{tot}-1$) the chunk is transferred  (Lines 1$-$5). 
Moreover, the chunks are assigned to different CUDA streams, which overlaps operations such as kernel execution and data transfer.
Prior to kernel execution, the chunk reads shared regions and then writes shared regions for reuse by the subsequent chunk (Line 6).
Subsequently, each chunk should be processed by a $k_{on}$-step kernel for $k^{'}_{off}/k_{on}$ times.
Before each invocation of the kernel, the computation area of the chunk must be adjusted because outer halo regions become irrelevant over time (Lines 7-10).
Similar to Line 3, if $k^{'}_{off}$ is indivisible by $/k_{on}$, the last time of execution involves applying a residual number of on-chip time steps to the chunk (Lines 11$-$14).  
After being processed for $k^{'}_{off}$ TB steps, the chunk is transferred back to the CPU to update the data (Line~15). 

Nevertheless, the reuse of intermediate results in the off-chip memory hinders the GPU kernels to exploit on-chip data reuse. To address the issue, we deliberately introduce redundant computation at the off-chip memory level, which decouples kernel execution from off-chip region sharing. 

\subsection{Redundant Computation in Off-Chip Memory}
\begin{figure}[t!]
    \centering
    \includegraphics[width=\linewidth]{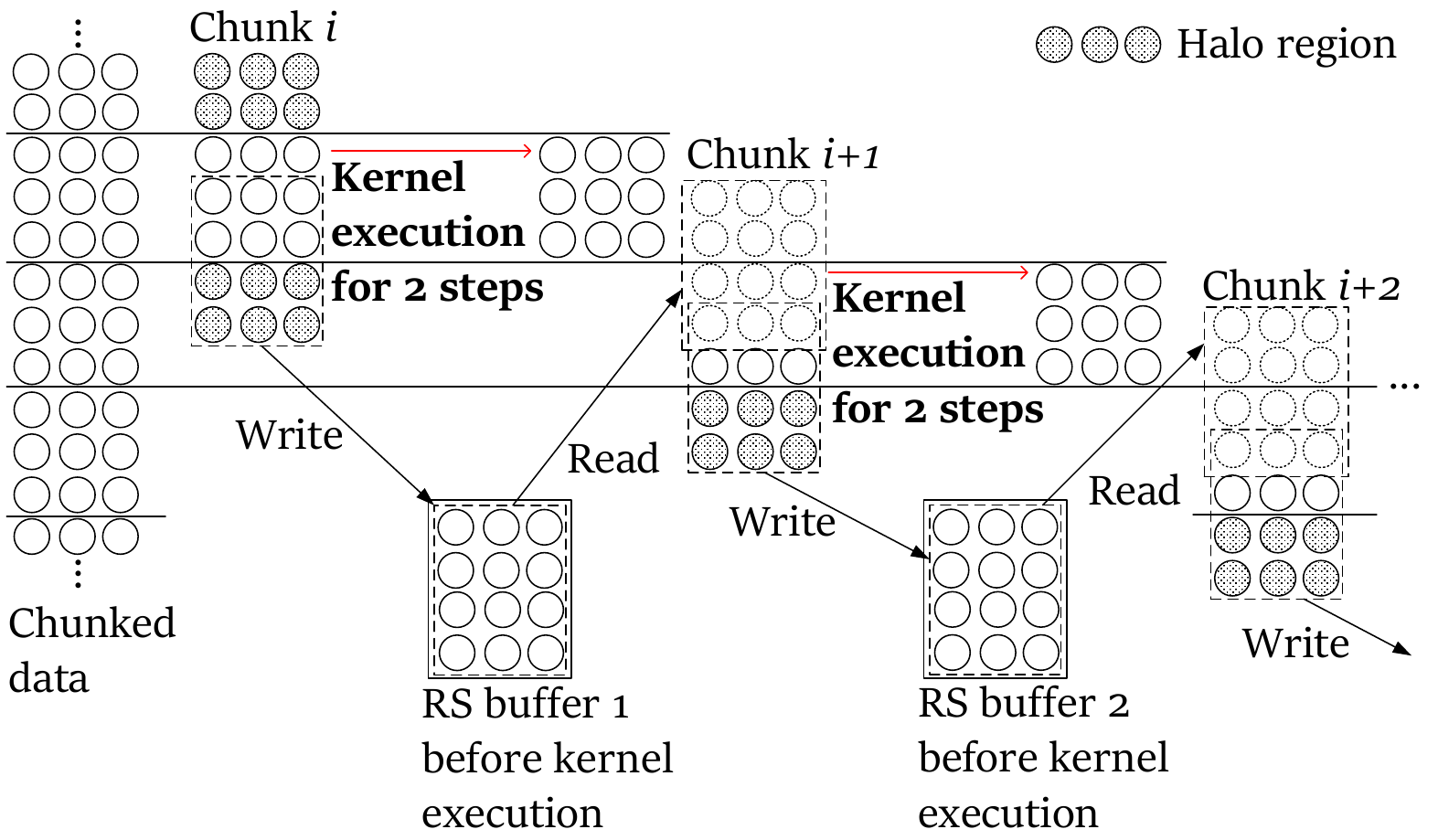}
    \caption{Introduction of redundant computation to resolve conflict between on- and off-chip data reuse. Note that ``RS'' represents ``region-sharing.''}
    \label{fig:redcomp}
\end{figure}
To resolve the conflict between on- and off-chip data reuse, we intentionally introduce redundant computation in the off-chip memory. 
As previously mentioned, existing region-sharing scheme conducts reads and writes of intermediate results in the overlapping regions between time steps, resulting in that each GPU kernel can merely execute for a single step.
In contrast, prior to each kernel invocation, SO2DR allows shared regions to be read and written for adjacent chunks to form an entirety of compute area.
As long as the entire compute area is formed, Each GPU kernel can execute for multiple time steps without interruption. 
In doing so, GPU kernels can leverage data reuse in the on-chip memory to improve performance (Fig.~\ref{fig:redcomp}).

Nonetheless, we must admit that SO2DR involves extra amounts of computation that pertain in the overlapping regions, hence redundant computation. However, such a side effect is easily outweighed by the performance benefit of exploiting on-chip data reuse, which will be demonstrated in the subsequent section.

\subsection{Run-time Parameter Selection}
\label{subsec:runtime}
\begin{table}[tb!]
\caption{Run-time configurations.}
    \centering
    \begin{tabular}{ll}
    \hline
    Variable & Description \\\hline
    $N_{a}$ & No. of arrays \\ 
    $b_{elem}$ & Data size of each array element \\ 
    $dim$ & No. of dimensions of an array \\ 
    $sz$ & Size along each dimension of an array\\
    $r$ & Stencil radius\\
    $d$ & No. of chunks \\
    $S_{tot}$ & Total time steps\\
    $S_{TB}$ & TB time steps\\
    $N_{strm}$ & No. of CUDA streams\\
    $C_{dmem}$ & Capacity of device memory\\
    $BW_{dmem}$ & Bandwidth of device memory\\
    $BW_{intc}$ & Bandwidth of interconnect between CPU and GPU\\\hline
    \end{tabular}
    \label{tab:conf}
\end{table}
The performance of GPU-based out-of-core stencil computation significantly depends on the selection of run-time parameters according to both the software (i.e., the stencil code) and hardware specifications. To this end, we offer a heuristic to select run-time parameters as follows.

The selection task is modeled with variables given in Table~I. Given the stencil code and experimental machine, \textit{the heuristic ensures a large ratio of execution time to data transfer time}:
 \begin{eqnarray*}
   \begin{array}{ll}
    \mathrm{satisfy}\hspace*{2mm}\frac{(D_{chk}+W_{halo}\times{S_{TB}})\times{n_{a}}}{BW_{dmem}}>\frac{D_{chk}\times{(n_{a}-1)}}{BW_{intc}},&\\
\mathrm{subject~to}\hspace*{2mm}(D_{chk}+W_{halo} \times S_{TB}) \times N_{strm} \leq \frac{C_{dmem}}{b_{elem}}, &\\
\hspace*{20mm}W_{halo}\times{S_{TB}} \leq D_{chk}, d>N_{strm},&\\
\mathrm{where}\hspace*{2mm}D_{chk}=\frac{sz(sz+2r)^{dim-1}}{d},W_{halo}=2r(sz+2r)^{dim-1}.&\\\end{array}
 \end{eqnarray*}

Here, in addition to the memory constraint, the working space allocated for halo regions ($W_{halo}\times{S_{TB}}$) cannot be larger than a chunk, otherwise the chunk fails to have a sufficient amount of data for region sharing.
Moreover, the number of chunks ($d$) must be larger than the number of CUDA streams ($N_{strm}$) to prevent streams from being idle. 
Lastly, $N_{strm}$ is fixed at three to efficiently overlap bidirectional data transfers as well as GPU execution (i.e., double buffering). Utilizing more than three CUDA streams is not considered capable of enhancing performance for large stencil codes~\cite{van2022generating}. 
Note that although the heuristic helps reduce the search space, it may select parameters that are feasible but not optimal. 
Therefore, we examine several feasible combinations of $d$ and $S_{TB}$ in the experiments, which will be presented and meticulously analyzed in the subsequent section.

\section{Evaluation}
\label{sec:expr}
In this section, experimental results are provided and analyzed to evaluate the effectiveness of SO2DR. 
\begin{table}[tb!]
\caption{Experimental machine.}
\begin{center}
\begin{tabular}{ll}
\hline
CPU & Intel Core i9-11900K\\
Host memory & 32~GB\\
GPU & NVIDIA GeForce RTX 3080\\
Device memory & 10~GB\\
PCIe & gen~3.0 $\times$16\\
Ubuntu & 18.04\\
CUDA & 10.2\\\hline
\end{tabular}
\label{tab1}
\end{center}
\end{table}

\begin{table}[tb!]
\caption{Benchmark stencil instances.}
\begin{center}
\begin{tabular}{lll}
\hline
Code        & box2d$x$r, $x\in$\{1,2,3,4\}& gradient2d\\
Access pattern &  (2$x$+1)$^2$ points & 5 points\\
Arithmetic intensity& \multirow{2}{*}{$2\times(2x+1)^2-1$}& \multirow{2}{*}{19}\\
(FLOPS/element)&  & \\
Value type &  \multicolumn{2}{c}{ single-precision floating-point}\\
Total size (in-core) & \multicolumn{2}{c}{12,800$\times$12,800 (1.2~GB)}\\
Total size (out-of-core) & \multicolumn{2}{c}{38,400$\times$38,400 (11.0~GB)}\\\hline
\end{tabular}
\label{tab1}
\end{center}
\end{table}
\begin{figure*}[tb!]
    \centering
    \includegraphics[width=\textwidth]{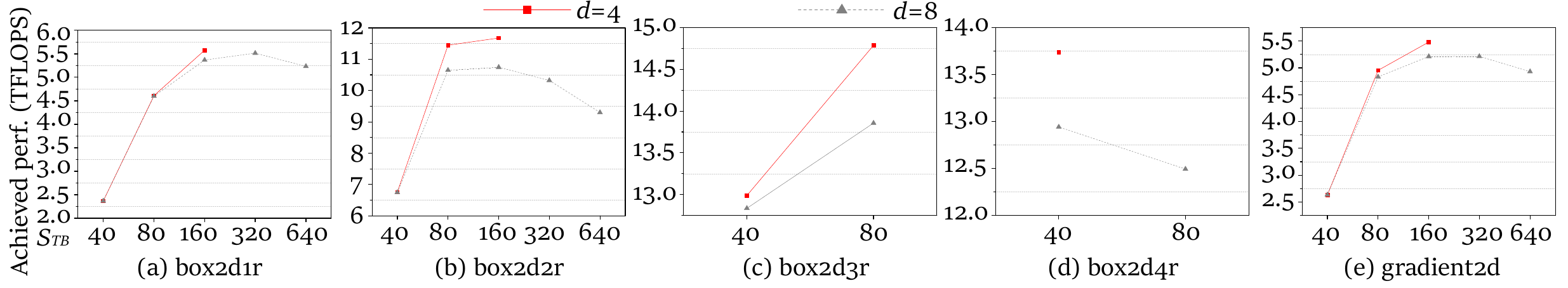}
    \caption{Performance achieved by using various run-time configurations.}
    \label{fig:tune}
\end{figure*}
\subsection{Experiental Setup}
A series of experiments were conducted on the experimental machine with the details given in Table~II.
The stencil computations used in the experiments are described in Table~III.

The proposed SO2DR implements region sharing with redundant computation at the off-chip memory level, and utilizes four-step GPU kernels.
Moreover, SO2DR was compared to two codes. The first is also an out-of-core code that employs the reuse of intermediate computation results between time steps~\cite{jin13cluster}. As previously mentioned, such a code can merely utilize single-step GPU kernels, denoted as the result-reuse code (ResReu). 
The second is an in-core code that utilizes four-step kernels as the proposed implementation does. The in-core code can only handle data that can fit in the GPU memory and is used to evaluate the performance impact of the out-of-core behavior of SO2DR. 
All GPU kernels, including the single-step kernels, were generated by the AN5D framework~\cite{matsumura2020an5d}. 

As for the run-time configurations, several candidate sets of run-time parameters were found using the model given in Section~\ref{subsec:runtime} (specifically, the number of chunks $d\in\{4,8\}$ and that of TB time steps $S_{TB}\in\{40,80,160,320,640\}$). These candidate configurations will be examined in terms of their performance impacts in the subsequent section.  
All the codes ran for a total number of 640 time steps. Moreover, three CUDA streams are utilized to overlap the operations on different chunks.
Furthermore, in addition to the out-of-core data (11.0~GB), we prepared the in-core data (1.2~GB) to compare in-core and out-of-core codes.

\subsection{Performance Evaluation for Candidate Run-time Configurations}

In this section, results obtained by executing SO2DR with out-of-core data (11.0~GB) and various combinations of run-time parameters~(Fig.~\ref{fig:tune}).
According to the results, a small number of chunks is favorable because it reduces transfer operations between the CPU and GPU. 
Moreover, a large ratio of kernel-execution time to data-transfer time is generally preferable but it is still limited by the memory-capacity constraint and a threshold of performance degradation. For instance, Figure~\ref{fig:tune}b showcases that for $d=8$, an $S_{TB}$ larger than 160 degrades the performance. In our experiments, while the ratio of kernel-execution time to data-transfer time is kept at a high level, a favorable ratio of the halo-region size to the chunk size is found to be less than 20\%. Note that this ratio may vary for different stencil codes and experimental environments.

Given the results, the configuration \{$d=4$,$S_{TB}=160$\} will be used for box2d\{1,2\}r and gradient2d,  \{$d=4$,$S_{TB}=80$\} for box2d3r, and {$d=4$,$S_{TB}=40$\} for box2d4r in the subsequent section.

\subsection{Comparison of Out-of-Core Codes in Terms of Performance}

\begin{figure}[tb]
    \centering
    \includegraphics[width=\linewidth]{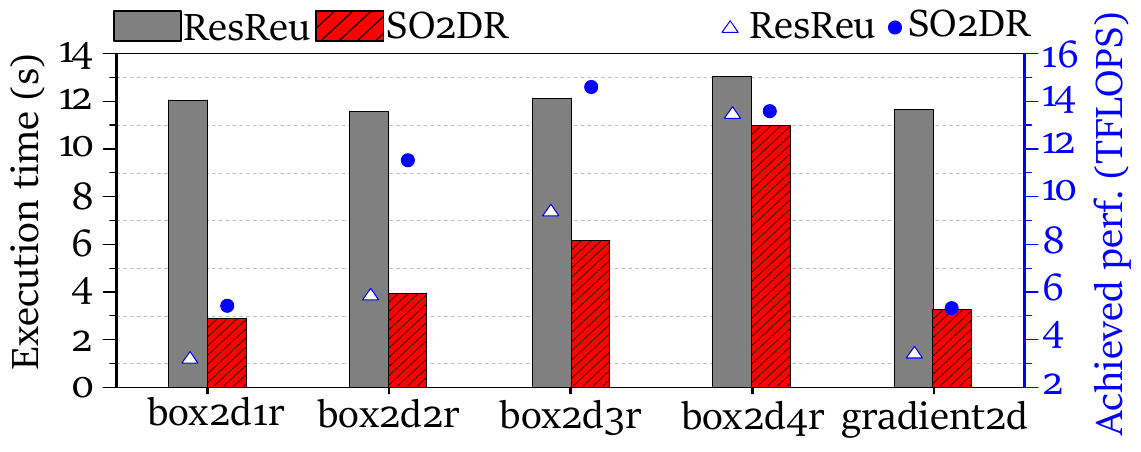}
    \caption{Comparison of out-of-core codes in terms of performance.}
    \label{fig:expr1}
\end{figure}
This section presents the results of comparing SO2DR to ResReu in terms of execution performance. The out-of-core data (11.0~GB) was used in the relevant experiments. 
As illustrated in Fig.~\ref{fig:expr1}, SO2DR outperforms the ResReu code for all evaluated stencil benchmarks.
Precisely, SO2DR achieves speedups of 4.22$\times$, 2.94$\times$, 1.97$\times$, 1.19$\times$, and 3.59$\times$ in comparison to ResReu for box2d1r, box2d2r, box2d3r, box2d4r, and gradient2d, respectively. 
The average speedup achieved by SO2DR is 2.78$\times$. 
Moreover, it is noteworthy that SO2DR performs best on moderate-order stencil instances (i.e., box2d\{1$-$3\}r and gradient2d where the stencil radii are 1, 2, 3, and 1, respectively).
For the high-order benchmark (box2d4r) whose radius is 4, the performance improvement achieved by SO2DR is limited to 19\%, indicating further kernel-execution optimizations are needed. 
\begin{figure}[tb]
    \centering
    \includegraphics[width=\linewidth]{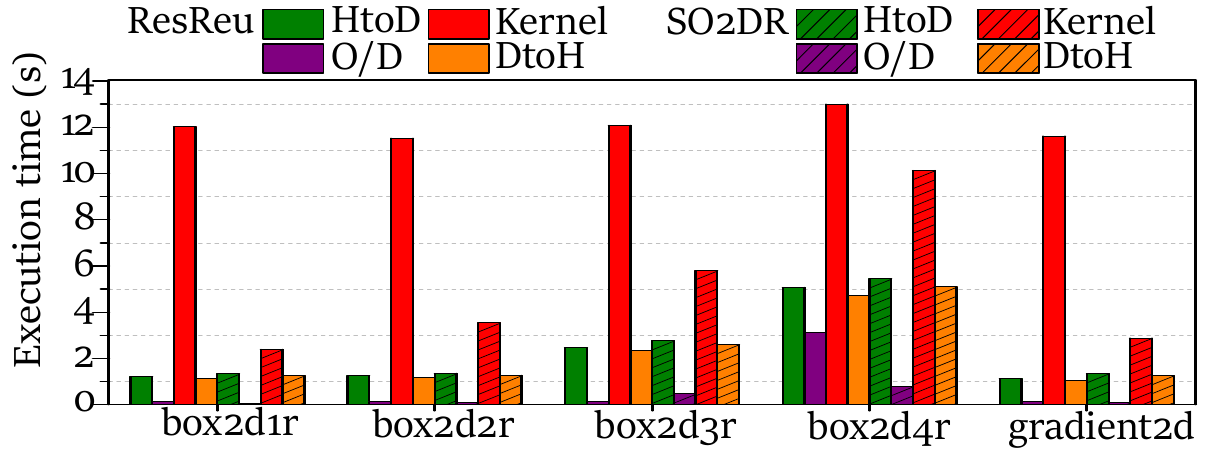}
    \caption{Breakdown analysis of performance achievement of proposed method. Note that ``HtoD,'' ``O/D,'' and ``DtoH'' represent ``host-to-device data transfer,'' ``on-device data copy,'' and ``device-to-host data transfer,'' respectively.}
    \label{fig:analysis1}
\end{figure}

Furthermore, Figure~\ref{fig:analysis1} presents the breakdown of execution time so as to analyze the high performance achieved by SO2DR.
The bottlenecks of both SO2DR and ResReu reside in kernel execution for all stencil benchmarks.
Nevertheless, SO2DR significantly reduces the kernel-execution time, hence enhancing the overall performance of out-of-core stencil computation.
On average, SO2DR reduces the execution time by 59\% on average, compared to ResReu. 

\begin{figure}[tb]
    \centering
    \includegraphics[width=\linewidth]{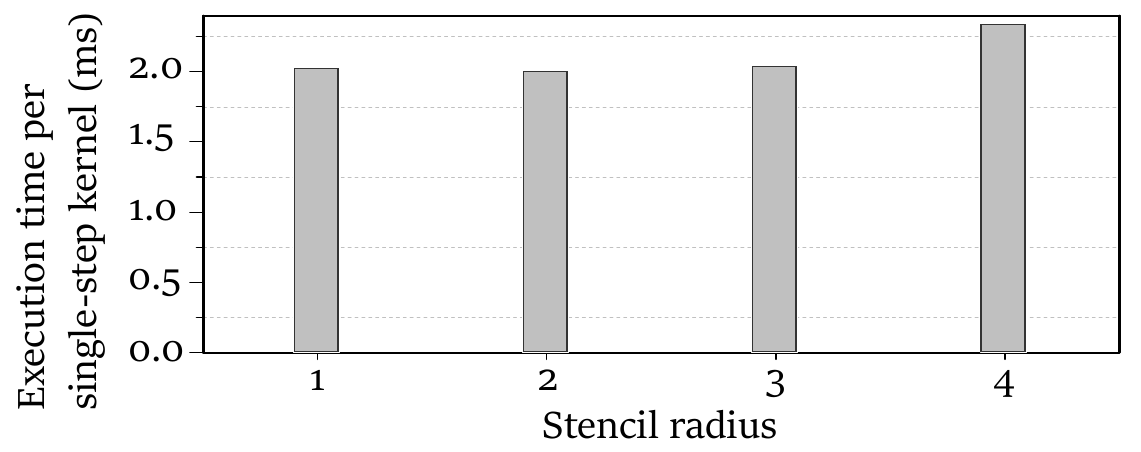}
    \caption{Average execution time per kernel measured for an in-core code which utilizes single-step kernels to handle box\{1$-$4\}r benchmarks.}
    \label{fig:analysis12}
\end{figure}

A less important but interesting observation is that the kernel-execution time of ResReu is almost the same for stencil benchmarks with various radii. 
For the cross validation of this observation, we measured the average execution time per kernel of an in-core code that utilizes single-step kernels and involves no CPU-GPU data transfers.   
The execution time per kernel is found definitely similar (Fig~\ref{fig:analysis12}).
This finding implies that the single-step kernels are inefficient, regardless of the stencil complexity. 

\subsection{Comparison of In-core and Out-of-Core Codes in Terms of Performance}
\begin{figure}[tb!]
    \centering
    \includegraphics[width=\linewidth]{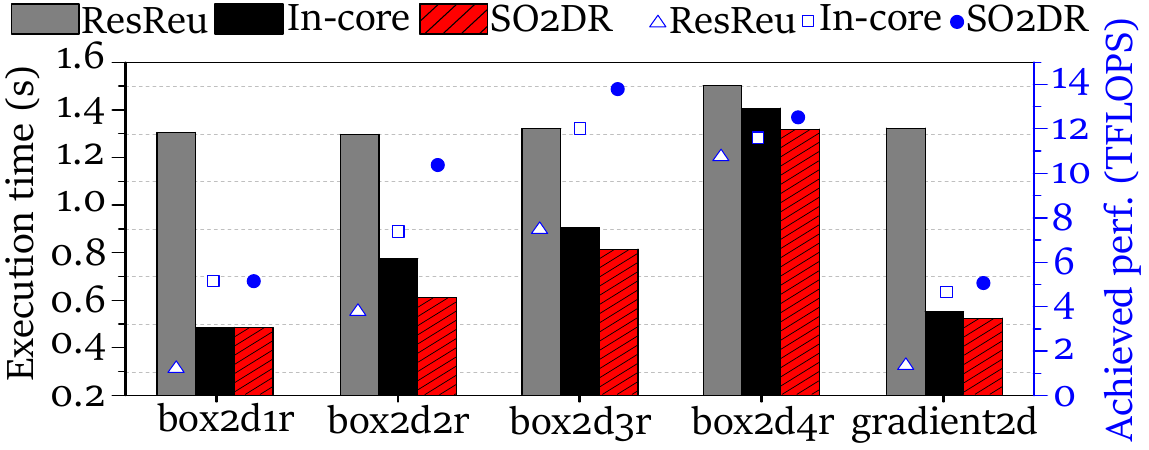}
    \caption{Comparison of incore and two out-of-core codes in terms of performance.}
    \label{fig:expr2}
\end{figure}
Moreover, we executed both the in-core and two out-of-core codes on the in-core dataset (1.2~GB), aimed at assessing the performance impact of the out-of-core behavior. 
The entirety of the in-core dataset fits within the device memory, allowing seamless processing by the in-core code. 
Moreover, because in-core codes requires two data transfer operations, i.e., the initial CPU-to-GPU transfer and the GPU-to-CPU transfer after all computations are finished, the data-transfer time is excluded from the performance evaluation.
As for the out-of-core codes, we followed the same chunk-based data decomposition strategy employed in previous experiments. 

As shown in Fig.~\ref{fig:expr2}, the the factors of performance degradation caused by ResReu are 105\%, 81\%, and 13\% for box2d\{2$-$4\}r, respectively.
Surprisingly, SO2DR even outperformed the in-core code, achieving speedups of $1.40\times$, $1.15\times$, $1.08\times$, and $1.08\times$, for box2d\{2$-$4\} and gradient2d, respectively. For box2d1r, SO2DR achieves the same performance as the in-core code does.
On average, SO2DR achieves a speedup of $1.14\times$, compared to the in-core code.
Such a performance achievement is attributed to the fact that SO2DR utilizes multiple operation streams which provide an opportunity of overlapping GPU kernels with each other.
Although ResReu also leverages multiple streams, the method suffers from long execution time of single-step kernels, which limits the overall performance.

\begin{figure}[tb]
    \centering
    \includegraphics[width=\linewidth]{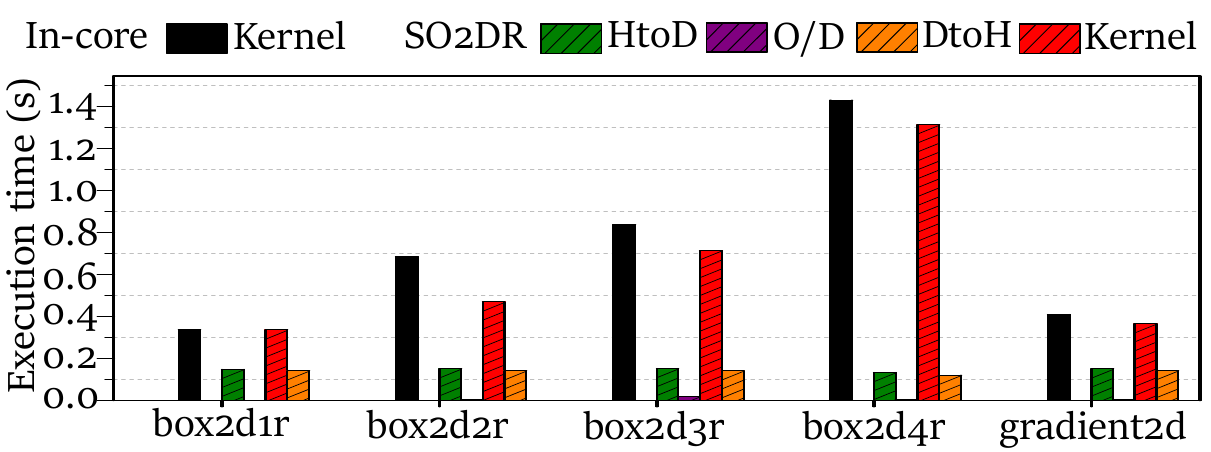}
    \caption{Breakdown analysis of performance degradation of proposed out-of-core method in comparison to in-core code. Note that ``HtoD,'' ``O/D,'' and ``DtoH'' represent ``host-to-device data transfer,'' ``on-device data copy,'' and ``device-to-host data transfer,'' respectively.}
    \label{fig:analysis2}
\end{figure}


Figure~\ref{fig:analysis2} presents the breakdown of execution time of SO2DR and the in-core code. 
Both codes are compute-bound and the overall performance is determined by the kernel-execution time.
SO2DR and the in-core method both implement four-step kernels but SO2DR exhibits improved kernel-execution time, thanks to the exploitation of multiple operation streams.

\section{Related Work}
\label{sec:rel}

Reguly $et~al.$~\cite{reguly2017loop} and Siklosi $et~al.$~\cite{siklosi2018heterogeneous} extended stencil applications to large systems with multiple compute nodes. Their efforts are majorly invested to distribute balanced workloads to compute nodes by meticulously decomposing and scheduling the loop executions of multi-stencil tasks. Likewise, Barreiros~$et~al.$~\cite{barreiros2022efficient} proposed a novel cost-aware data partitioning strategy to accelerate image-analysis stencil tasks running in CPU-GPU systems. The primary focus is to address the load-balancing problem. Kernel optimizations are not mentioned in these studies.

Mudalige $et~al.$~\cite{mudalige2019large} efficiently ported a large-scale extensively-utilized legacy stencil application to present massively parallel hardware. They overcame the challenges of data organization and transfer by re-engineering the application with techniques such as loop scheduling. Nonetheless, tiling in the on-chip memory of accelerators is not mentioned. 

Sun $et~al.$~\cite{sun2020burstz,sun2022burstz+} presented compression-based approaches to address the performance bottleneck imposed by data transfer during executing stencil tasks in accelerator-based (e.g., CPU-GPU and CPU-FPGA) systems. Their work can be leveraged in combination with ours to further enhance the performance. 

Qu $et~al.$~\cite{qu2021toward,qu2022exploiting} leveraged an asynchronous multi-core wavefront diamond tiling approach to optimize stencil computational kernels, enhancing data reuse during temporal blocking is being performed. 
Nonetheless, diamond tiling introduces complex dependencies between tiles (i.e., chunks), which hinders the process of streaming chunks to and from the GPU. To this end, our work utilizes parallelogram tiling which is efficient for GPU-based out-of-core stencil computation.

Sioutas $et~al.$~\cite{sioutas2020schedule} proposed a novel optimization to schedule the pipelines of image-processing stencil applications which are running on CUDA-bases GPU systems.
Li $et~al.$~\cite{li2022revisiting} proposed a novel approach to split a parallelogram tile into multiple phases that can be concurrently executed with those of the neighboring tiles. Both studies do not consider the out-of-core scenario.

Van Beurden $et~al.$~\cite{van2022generating} systematically investigated factors that influence the performance of GPU-based out-of-core stencil codes by deliberately varying the ratio of computation to communication. They also studied factors that determine the optimal number of CUDA streams. Nevertheless, tiling at the on-chip memory level is not considered.  

\section{Conclusion and Future Work}
\label{sec:conc}
In this work, we propose SO3DR, a synergy between on- and off-chip data reuse to significantly enhance the performance of GPU-based out-of-core computation.
SO2DR decouples kernel execution from region sharing by intentionally introducing redundant computation in the off-chip memory. In doing so, the functionality of region sharing in the off-chip memory is preserved to eliminate redundant data transfer between the CPU and GPU. Furthermore, data reuse/tiling in the on-chip memory is enabled to considerably improve kernel execution.
Experimental results demonstrate the effectiveness of SO2DR on five representative stencil benchmarks, compared to an out-of-core competitor which can eliminate both redundant data transfer and redundant computation.

As for future work, we first consider extending this work to multi-stencil codes and more distributed systems.
We also plan to refine the performance model which can be used to automatically select the optimization target between kernel execution and data transfer. 

\bibliographystyle{ieeetr}
\bibliography{ref}

\end{document}